\documentclass[12pt,preprint]{aastex}

\newcommand{\halpha}{H$\alpha$\ }
\newcommand{\vsini}{v {\rm sin} i}

\slugcomment{Accepted for publication in ApJ Letters}

\shorttitle{APOGEE Discovers 2 Sig Ori E Stars}
\shortauthors{Eikenberry, et al.}

\begin{document}


\title{Discovery of Two Rare Rigidly-Rotating Magnetosphere Stars in the APOGEE Survey}


\author{Stephen S. Eikenberry\altaffilmark{1}, S. Drew Chojnowski
  \altaffilmark{2}, John Wisniewski\altaffilmark{3}, Steven
  R. Majewski\altaffilmark{2}, Matthew Shetrone\altaffilmark{4}, David
  G. Whelan\altaffilmark{2,5}, Dmitry Bizyaev\altaffilmark{6,7},
  H. Jacob Borish\altaffilmark{2}, James R.A. Davenport
  \altaffilmark{8}, Garrett Ebelke\altaffilmark{6,7}, Diane
  Feuillet\altaffilmark{7}, Peter M. Frinchaboy\altaffilmark{9}, Alan
  Garner \altaffilmark{1}, Fred Hearty\altaffilmark{2}, Jon Holtzman
  \altaffilmark{7}, Zhi-Yun Li\altaffilmark{2},
  Sz.~M{\'e}sz{\'a}ros\altaffilmark{10,11}, David L. Nidever
  \altaffilmark{2,12}, Donald P. Schneider\altaffilmark{13}, Michael
  Skrutskie\altaffilmark{2}, John C. Wilson\altaffilmark{2}, Gail
  Zasowski\altaffilmark{14}}


\altaffiltext{1}{Department of Astronomy, University of Florida}
\altaffiltext{2}{Department of Astronomy, University of Virginia}
\altaffiltext{3}{Department of Astronomy, University of Oklahoma}
\altaffiltext{4}{University of Texas, McDonald Observatory}
\altaffiltext{5}{Department of Physics \& Astronomy, Hampden-Sydney
  College} 
\altaffiltext{6}{Apache Point Observatory}
\altaffiltext{7}{Department of Astronomy, New Mexico State
  University} 
\altaffiltext{8}{Department of Astronomy, University of Washington}
\altaffiltext{9}{Department of Physics \& Astronomy, Texas Christian University}
\altaffiltext{10}{Instituto de Astrof{\'{\i}}sica de Canarias (IAC), E-38200 La Laguna, Tenerife, Spain}
\altaffiltext{11}{Departamento de Astrof{\'{\i}}sica, Universidad de La Laguna (ULL), E-38206 la Laguna, Tenerife,
Spain}
\altaffiltext{12}{Department of Astronomy, University of
  Michigan} 
\altaffiltext{13}{Department of Astronomy\& Astrophysics,
  The Pennsylvania State University} 
\altaffiltext{14}{Department of Physics \& Astronomy, Johns Hopkins University}


\begin{abstract}
The Apache Point Observatory Galactic Evolution Experiment (APOGEE) -
one of the Sloan Digital Sky Survey III programs -- is using
near-infrared spectra of $\sim 100,000$ red giant branch star
candidates to study the structure of the Milky Way. In the course of
the survey, APOGEE also acquires spectra of hot field stars to serve
as telluric calibrators for the primary science targets. We report the
serendipitous discovery of two rare, fast-rotating B stars of the
$\sigma$ Ori E type among those blue field stars observed during the
first year of APOGEE operations. Both of the discovered stars display
the spectroscopic signatures of the rigidly rotating magnetospheres
(RRM) common to this class of highly-magnetized ($B \sim 10$
kiloGauss) stars, increasing the number of known RRM stars by $\sim 10
\%$.  One (HD 345439) is a main-sequence B star with unusually strong
He absorption (similar to $\sigma$ Ori E), while the other (HD 23478)
fits a ``He-normal'' B3IV classification.  We combine the APOGEE
discovery spectra with other optical and near-infrared spectra of
these two stars, and of $\sigma$ Ori E itself, to show how
near-infrared spectroscopy can be a uniquely powerful tool for
discovering more of these rare objects, which may show little/no RRM
signatures in their optical spectra. We discuss the potential for
further discovery of $\sigma$ Ori E type stars, as well as the
implications of our discoveries for the population of these objects
and insights into their origin and evolution.

\end{abstract}





\section{Introduction}

$\sigma$ Orionis E is the archetype of an unusual and rare class of
helium-strong main sequence B stars \citep{Gray2009}, characterized by
extremely large magnetic fields and fast rotation.  $\sigma$ Ori E
itself has a measured longitudinal magnetic field varying with an
amplitude of $B_l \sim 2-3$ kG with an inferred polar magnetic
strength of $\sim 10$ kG \citep{Townsend05, Kochukhov11, Oksala2012} a
rotational velocity of $\vsini = 160 \ {\rm km \ s^{-1}}$, and a
rotational period of $1.19 \ {\rm d}$ \citep{Townsend05}.  The high
magnetic field of the star is thought to form a Rigidly Rotating
Magnetosphere (RRM) \citep{Townsend05} which traps circumstellar
material in two co-rotating clouds at a distance of several stellar
radii beyond the photospheric surface, producing an extremely broad,
double-horned \halpha emission profile with velocity width $>1000
\ {\rm km \ s^{-1}}$ as well as periodic modulation of the star's
light curve. The magnetic field also appears to be responsible for the
enhanced He absorption via localized surface abundance anomalies it
creates in the star. A more recently-discovered star in the same
class, HR 7355 \citep{Riv2008, Riv2010, Oksala2010}, shows similarly He-strong
absorption and polar field strength ($B \sim 11-12$ kG) to $\sigma$
Ori E, and corresponding \halpha emission profiles with velocity
widths of $\sim 1300 \ {\rm km \ s^{-1}}$.  HR 7355 shows
exceptionally fast rotation ($P = 0.52$ d and $\vsini = 310 \ {\rm km
  \ s^{-1}}$); this star, along with HR5907 ($P = 0.51$d, $\vsini =
340 \ {\rm km \ s^{-1}}$), are the two fastest-rotating,
non-degenerate magnetic stars known -- in fact, their speeds approach
the rotational breakup velocity \citep{Riv2013, Grunhut}.  The
simultaneous presence of such extreme rotation and large magnetic
field is somewhat surprising for massive B stars, and for HR 7355 the
spindown timescale via magnetic braking should be much shorter than
its estimated age \citep{Riv2013, Mikulasek10}. Thus, these stars
provide a unique conundrum for theories of both star formation and
magnetic field evolution.  Increasing the known number of these
objects will allow us to understand how common this phenomenon is for
massive stars, and establish the range of properties they can exhibit.

In this paper, we present the discovery of two additional members of
this rare class of stars from the Sloan Digital Sky Survey III's
Apache Point Observatory Galactic Evolution Experiment
(SDSS-III/APOGEE) \citep{Gunn, Eisenstein11, Majewski2013,
  Wilson2012}.  APOGEE is a near-infrared (NIR), H-band ($1.51-1.70
\ \mu$m), high-resolution ($R \simeq 22,500$), spectroscopic survey
primarily targeting red giant stars in the Milky Way.  As part of
routine survey operation, APOGEE selects bright, hot, blue stars in
each survey field for telluric correction of the red giant spectra.
Below, we describe the serendipitous discovery of the two $\sigma$ Ori
E type stars among these APOGEE telluric standards, identified via the
unique magnetospheric signatures they produce in their Brackett series
emission profiles.  We next present optical spectra that confirm the
He-strong classification of one star and the apparent ``He-normal''
nature of the other, allow measurements of $\vsini$, and show evidence
of \halpha profiles matching the Brackett series RRM signatures. We
also present Triplespec NIR spectra of these two stars and $\sigma$
Ori E itself, which confirm the similarities among all three stars and
the ``smoking gun'' signature provided by the Brackett series line
profiles created by the rigidly rotating magnetospheres in these
stars.  We conclude by discussing the potential for further discovery
of $\sigma$ Ori E type stars during the APOGEE survey, and the
implications they will have for understanding the breadth of
characteristics in the population of these objects, their origin, and
their evolution.


\section{Observations and Spectral Analyses}

\subsection{APOGEE Near-IR Discovery Spectra}

To assess and remove telluric absorption features from APOGEE science
spectra, 35 of the 300 instrument fibers used in each observation are
used to observe hot, blue stars simultaneously with the normal science
targets \citep{Gail2013}. The first 18 telluric calibrators selected
are subject to a spatial constraint, whereby the APOGEE field in
question is divided into 18 equal-area zones and the bluest star (in
raw $J-K_s$ color, with $5 < {\rm H} < 11$ mag) in each zone is
selected as a target. The remaining 17 telluric calibrators are simply
the bluest stars anywhere in the field of view not previously
selected. Visual inspection of these spectra led to the identification
of numerous Be stars, which we placed into a numbered list (in order
of discovery) for further study. HD 23478 (ABE-075) was observed a
total of nine times, with apporximately 1-hr observation each time, on
the six different fiber plug plates designed for a special study of
the young open cluster IC348. HD 345439 (ABE-050), however, was only
observed on an APOGEE instrument commissioning fiber plate (a plate
designed for testing of sky and telluric removal). While this plate
was observed twice, the second observation produced poor
S/N. Fortunately the first observation achieved good S/N ($\sim 70$
per resolution element in the continuum).

In Figure 1, we present the APOGEE NIR discovery spectra for the two
newly-identified $\sigma$ Ori E stars -- labeled ABE-050 (HD 345439)
and ABE-075 (HD 23478) in the APOGEE catalog of Be stars.  In the
APOGEE bandpass, we can see prominent Brackett series emission lines
(Br11-Br20) from both stars, each with a characteristic double-horned
profile.  In Figure 2, we display a detailed view of the Br11 line
profiles from each star -- as may be seen, the profile peak
separations approach $\sim 1000-1100 \ {\rm km \ s^{-1}}$, which is
$\sim 2$ times greater than the largest linewidths seen from more
``normal'' Be stars observed by APOGEE \citep{Drew_paper}.  This type
of profile and peak separation is typical of the RRM feature of
$\sigma$ Ori E stars, and is strong evidence that HD 345439 and HD
23478 are likely to be rapidly-rotating, highly magnetized stars.  In
short, the APOGEE infrared color selection for telluric standards
(which selects stars with colors roughly similar to B stars) and
APOGEE spectra alone make these stars strong candidates for the
$\sigma$ Ori E type classification.

\begin{figure}
\epsscale{0.6}
\plotone{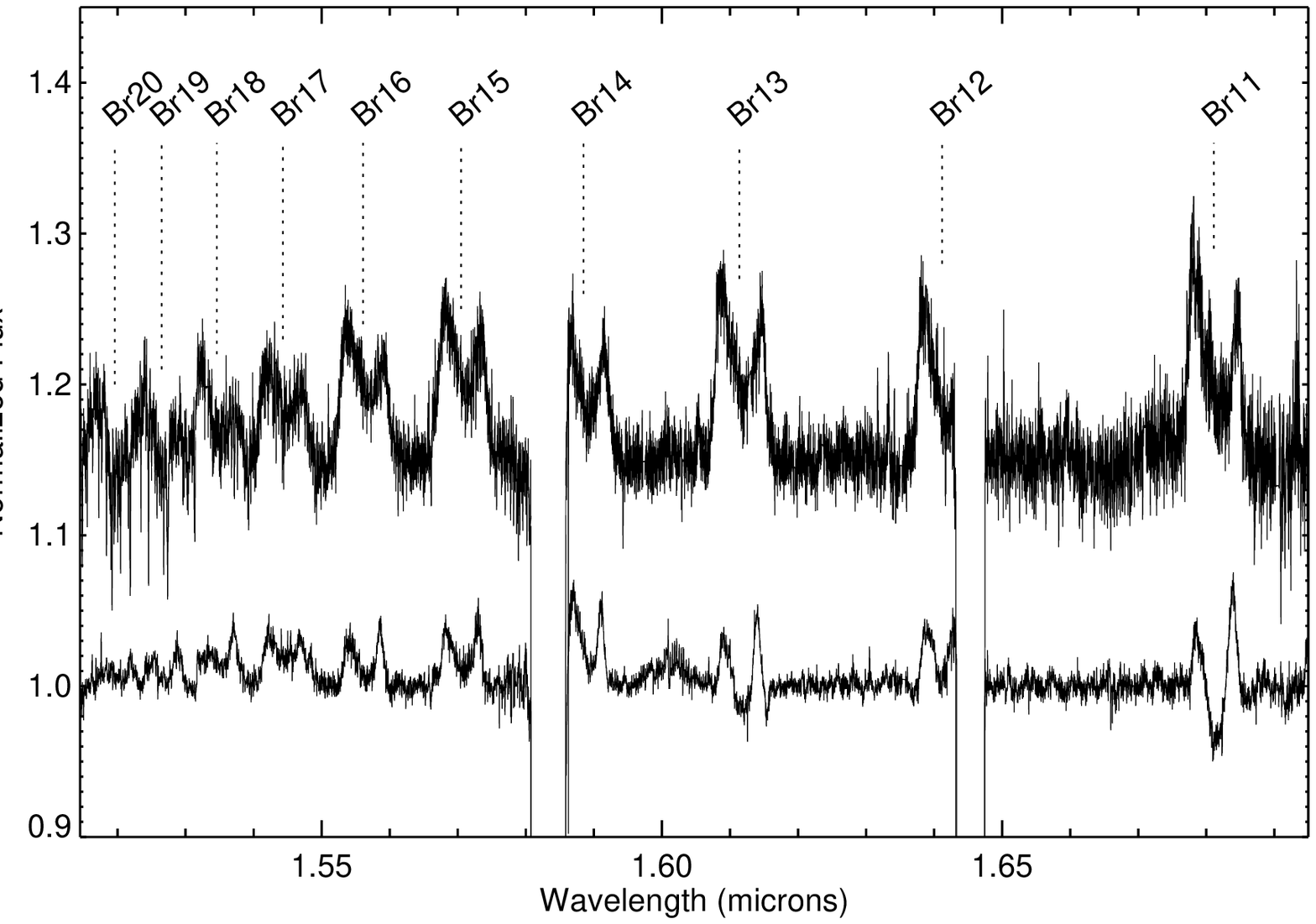}
\plotone{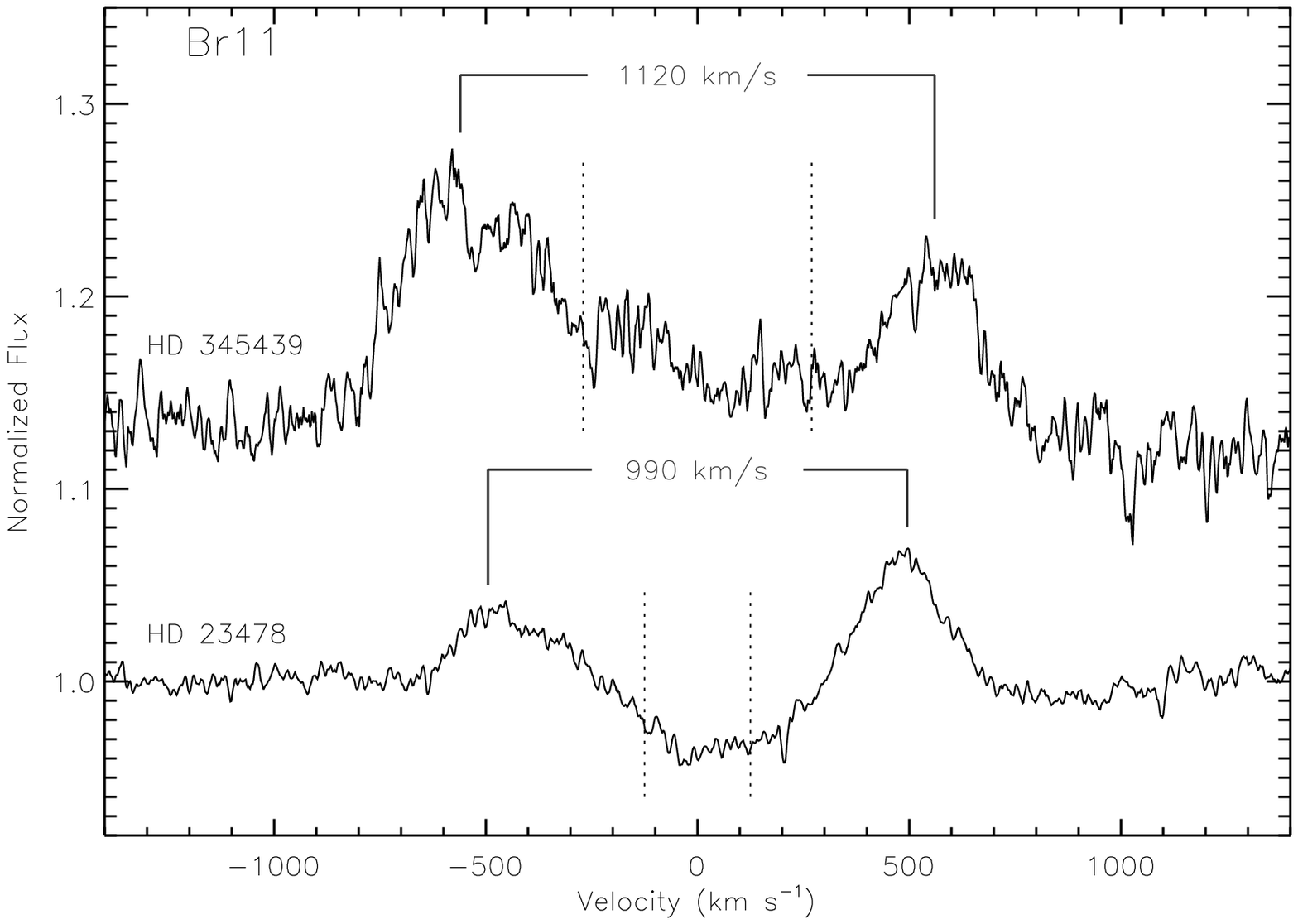}
\caption{(Top) APOGEE discovery spectra of the emission-line character
  of the stars HD 345439 (upper) and HD 23478 (lower).  The pronounced
  double-horned Brackett series emission lines dominate the $H$-band
  spectrum of these stars.  The gaps near $1.583 \mu$m and $1.647
  \mu$m are due to the inter-detector spacing in the APOGEE detector
  focal plane. (Bottom) Expanded view of the Br11 ($1.681 \mu$m)
  profiles for HD 345439 (top) and HD 23478 (bottom). The
  double-horned velocity profile with peak separations of $\sim 1000
  {\rm km \ s^{-1}}$, significantly exceeding the range of $\pm
  \vsini$ (indicated by the dotted vertical lines), is characteristic
  of the Rigidly Rotating Magnetosphere (RRM) stars of the $\sigma$
  Ori E class. The spectra have been rebinned by a factor of four from
  the native APOGEE resolution for an improved signal-to-noise ratio
  for these broad features.  The spectrum of HD 345439 has been offset
  by 0.1 in intensity for clarity.}
\end{figure} 

\begin{figure}
\epsscale{0.7}
\plotone{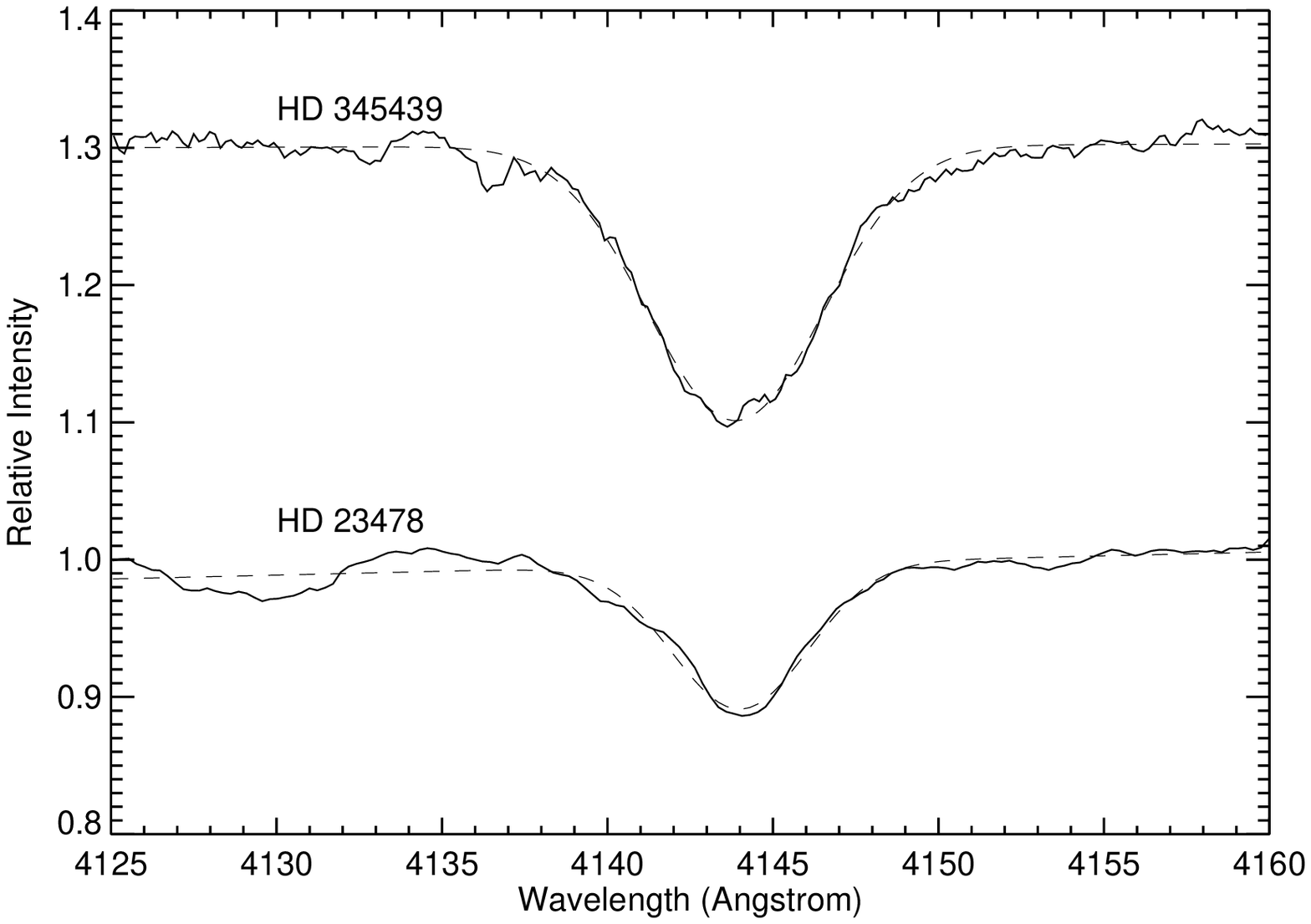}
\caption{Line profile fits to HeI absorption lines used to determine
  $\vsini$ .  We used simple Gaussian absorption profiles, which seem to
  provide accurate fits for both stars' broad absorption lines.  We use multiple lines for each star - the plot above shows HeI ${\rm \lambda 4144 \ \AA}$}
\end{figure}

\subsection{Optical and Triplespec Near-IR Spectra}

Based on this initial identification, we obtained followup optical
spectra of HD 345439 and HD 23478, as well as $1-2.5 \ \mu$m NIR
spectra of both stars and $\sigma$ Ori E itself to confirm the
identification.  In the optical, we observed both stars using the High
Resolution Spectrograph (HRS) \citep{Tull98} on the Hobby-Eberly
Telescope (HET, \citet{Ramsey98}) as part of queue-scheduled observing
in the lower priority band \citep{Shetrone2007} at $R = 18000$ with
the cross disperser set to achieve spectra from $ 3910 - 4880$ \AA\ on
the blue detector and from $ 4990 - 6820$ \AA\ on the red detector.  We
reduced the spectra with IRAF\footnote{IRAF (Image Reduction and
  Analysis Facility) is distributed by the National Optical Astronomy
  Observatories, which are operated by the Association of Universities
  for Research in Astronomy, Inc., under contract with the National
  Science Foundation.} ECHELLE tasks, using the standard IRAF tasks
for overscan removal, bias subtraction, flat fielding, scattered light
removal and wavelength calibration.  We present the resulting spectra
in Figure 3.

At first glance, the optical spectra of both stars - especially the
region blueward of \halpha -- resemble those of typical B-stars, with
strong Balmer absorption lines as well as prominent He absorption
features.  However, the absorption lines show very broad symmetric
profiles that indicate high rotational velocities, and for HD 345439
the HeI lines are unusually strong.  Comparing its spectrum to the
typical B stars in \cite{Gray2009}, the presence and strength of the
HeI lines constrain HD 345439 to be in the spectral range of O9 to B3.
The absence of HeII ${\rm 4686 \ \AA}$ constrains both stars to be
later than B0, and HD 345439 closely matches B1V or B2V stars in the
strength of SiII ${\rm \lambda \lambda \ 4128 - 4130 \ \AA}$ and MgII
${\rm \lambda \ 4481 \ \AA}$, which are key indicators in this
sequence \citep{WalbornFitzpatrick}.  Furthermore, the strength of HeI
${\rm 4026 \ \AA}$, OII ${\rm 4070-4076 \ \AA}$, and the lack of
detectable emission from OII ${\rm 4348 \ \AA}$, OII ${\rm 4416
  \ \AA}$ and Si III ${\rm 4553 \ \AA}$ all confirm the main sequence
classification of this star. However, none of the ``normal'' stars in
\cite{Gray2009} possess the HeI absorption strength we see in HD
345439.  HD 23478, on the other hand, appears optically to be a fairly
``normal'' star at first glance, based on the blue part of its
spectrum, with much weaker He absorption than HD 345439.  This object
was previously classified as a B3IV star \citep{hiltner56, crawford58,
  walker63}, and this classification is confirmed in our spectrum by
the presence and strength of the MgII ${\rm 4481 \ \AA}$, CII ${\rm
  4267 \ \AA}$, and SiII ${\rm 4128-4130 \ \AA}$ absorption features.
However, the HeI ratio of ${\rm 4144 \ \AA / 4121 \ \AA}$ is $\ll 1.0$
in HD 23478 -- a peculiarity not seen in other stars, where the ratio
is typically $\ga 1.0$ \citep{WalbornFitzpatrick, Gray2009}.
Furthermore, the \halpha features in both HD 345439 and HD 23478
display the pronounced broad emission typical of RRM stars in both HD
23478 and HD 345439.

\begin{figure}
\epsscale{1.0}
\plotone{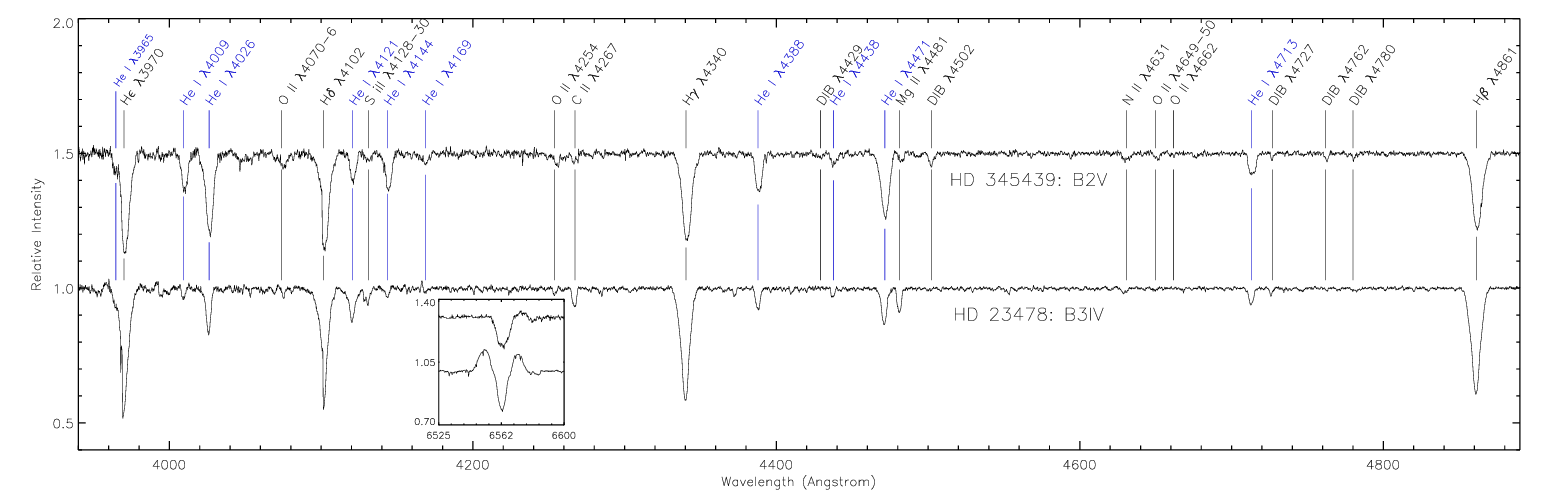}
\caption{HET optical spectra of HD 345439 and HD 23478, with HD 345439
  offset by 0.3 in relative intensity from HD 23478, and including key
  line identifications for spectral classification (including some
  Diffuse Interstellar Bands - DIBs).  Inset plot shows the region
  around \halpha for both stars.  The HeI features are generally
  stronger and broader in HD 345439 than in HD 23478.  HD 23478
  exhibits the broad \halpha RRM signatures at this observation epoch,
  while HD 345439 shows only a slight emission ``bump'' on the red
  side at this epoch.  These spectra are not contemporaneous with the
  NIR spectra from APOGEE nor Triplespec.}
\end{figure} 

We also obtained Triplespec \citep{Wilson2004} NIR spectra of HD
345439, HD 23478, and $\sigma$ Ori E itself (Figure 4), using the APO
3.5m. We used the 1$\farcs$1 slit, for $R \sim 3500$ spectra between
0.95-2.46 $\mu$m, and all data were acquired nodding in ABBA
mode. Eight 90 second integrations of HD 345439 and twenty 20 second
integrations of the A0V star HD 189690, used for telluric correction
\citep{Vacca03}, were obtained on 2012 September 3.  Eight 30 second
integrations of HD 23478 and four 30 second integrations of the A0V
star HR 1724 were performed on 2013 February 15.  Six 60 second
integrations of $\sigma$ Ori E and six 60 second integrations of the
A0V star HD 67015 were obtained on 2013 January 25.  These data were
reduced using Triplespectool, a modified version of Spextool developed
for use with the SpeX instrument at IRTF \citep{Cushing04}. The
strongest features of all three spectra are again the extremely broad
double-horned Brackett-series emission profiles that distinguish the
rigidly-rotating magnetospheres of these stars.  HD 345439 appears to
have the strongest RRM feature at these particular epochs of
observation, while HD 23478 and $\sigma$ Ori E are similar in emission
strengths/widths (with $\sigma$ Ori E being slightly more asymmetric
and with deeper absorption depth at these epochs).  Here again HD
23478 shows significantly weaker HeI absorption features (i.e., at
$1.70 \mu$m and at $2.11 \mu$m) than the other two stars.  The
persistence of this weak absorption at multiple epochs seems to
indicate that HD 23478 may truly be more ``He-normal'' than HD 345439
and the RRM archetype $\sigma$ Ori E - though more data are needed to
confirm this conclusion.

\begin{figure}
\epsscale{1.0}
\plotone{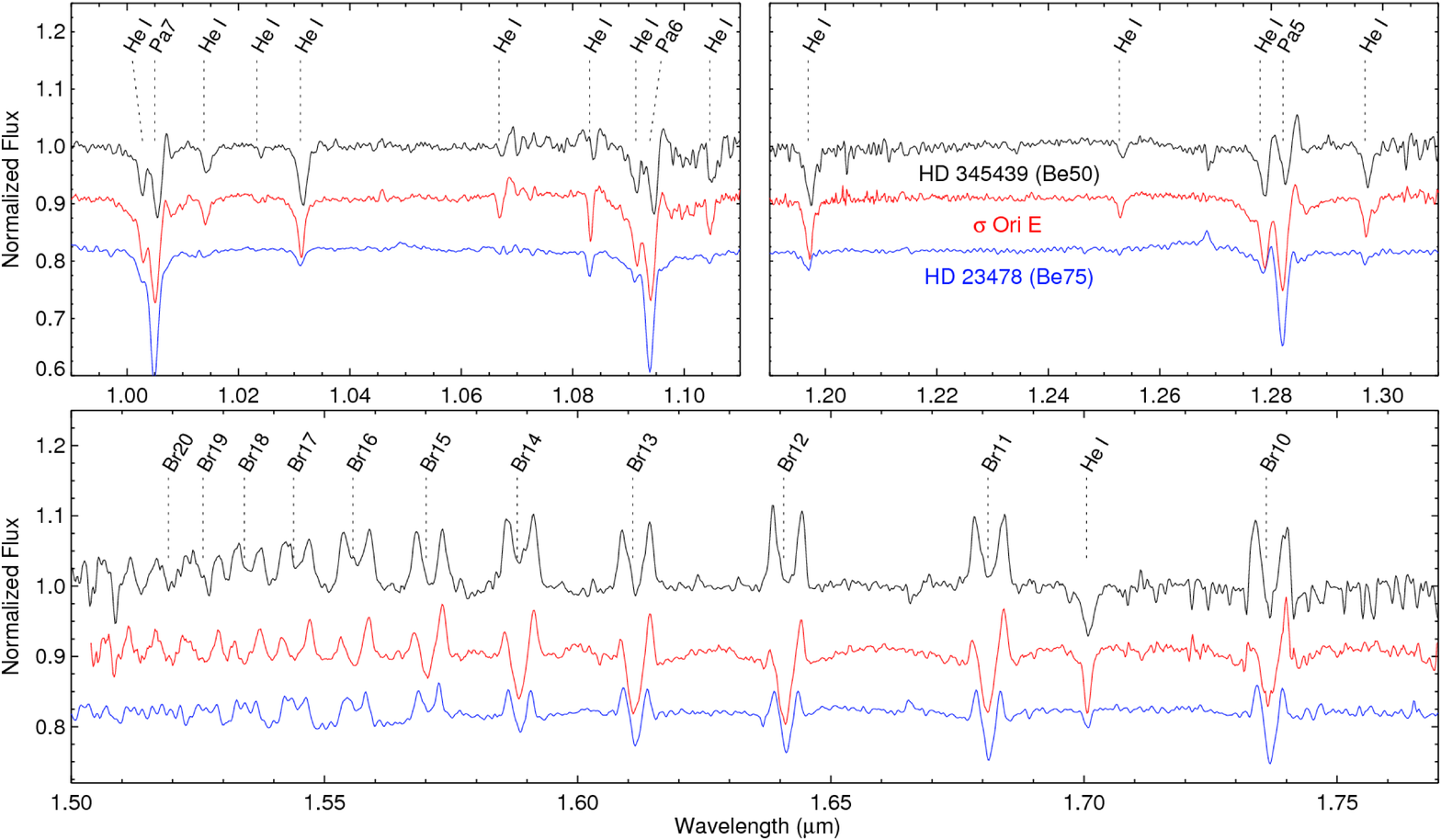}
\caption{Portions of the normalized Triplespec NIR spectra of HD
  345439 (black), HD 23478 (blue), and the RRM star archetype $\sigma$
  Ori E (red).  The RRM features are seen in all lines of the Brackett
  series. Also, HD 23478 is noticeably weaker in its HeI absorption
  features than the other two stars.}
\end{figure} 

To measure $\vsini$ for the stars, we fit line profiles for the HeI
absorption features at ${\rm 4026 \ \AA}$, ${\rm 4144 \ \AA}$, ${\rm
  4388 \ \AA}$ and ${\rm 4471 \ \AA}$, and account for non-rotational
effects according to the correction factors in \citet{Daflon2007},
which then lead to $\vsini$ values of $270 \pm 20 \ {\rm km \ s^{-1}}$
for HD 345439 and $125 \pm 20 \ {\rm km \ s^{-1}}$ for HD 23478.
These measurements are typical of the fast-rotating highly-magnetized
$\sigma$ Ori E stars -- the $\vsini$ for HD 23478 is very similar to
(albeit slightly slower than) the value for $\sigma$ Ori E itself,
while HD 345439 appears to be one of the fastest known rotators among
main sequence stars, surpassed only by the two recently discovered
$\sigma$ Ori E stars, HR 7355 and HR 5907.  Previous work
\citep{Jerzy} indicates a photometric period for HD 23478 of
$1.0499$d, which is also very similar to (and slightly faster than)
the measured $1.19$d rotation period of $\sigma$ Ori E.  These
rotational properties confirm our classification of HD 345439 and HD
23478 as $\sigma$ Ori E stars.

We measured the HeI equivalent widths for both stars and present them
along with values for typical B stars, as well as $\sigma$ Ori E, in
Figure 5. Again, this confirms that HD 345439 is typically
``He-strong'', as are other $\sigma$ Ori E stars, while HD 23478 may
be ``He-normal'', despite being a fast rotator with the clear
signature of a RRM.  We note that helium absorption strength can be
phase-dependent in the $\sigma$ Ori E stars, and the variations may be
large enough to mask their ``He-strong'' nature at some rotational
phases, so additional phase-sampled spectra are required to confirm
this conclusion.

\begin{figure}
\epsscale{1.0}
\plotone{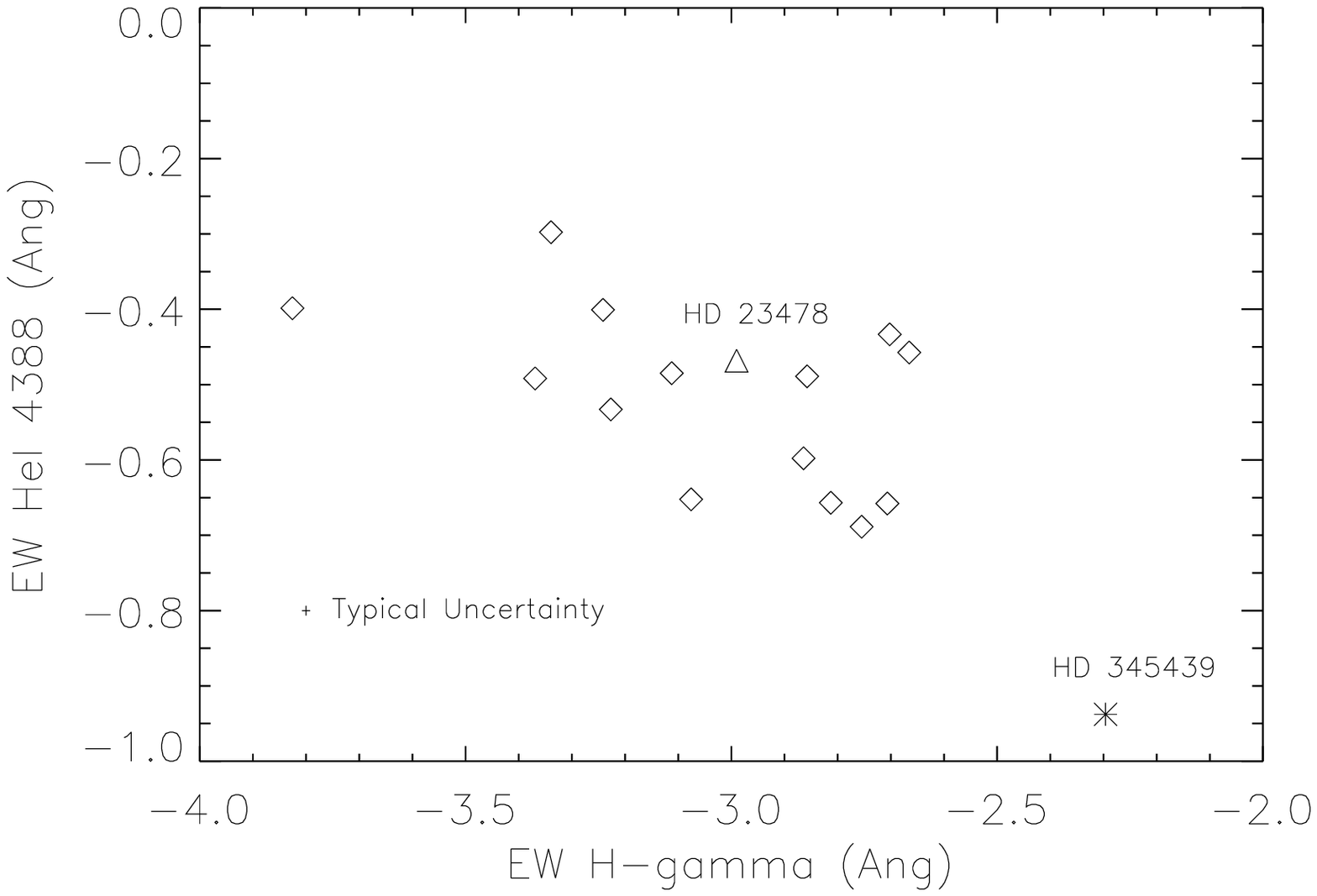}
\caption{HeI and ${\rm H \gamma}$ equivalent widths for the HET
  standard B stars (diamonds), as well as HD 345439 (asterisk) and HD
  23478 (triangle).  HD 345439 clearly is an outlier from the normal
  B-stars, while HD 23478 appears to be ``He-normal'' in these
  observations.}
\end{figure}




\section{Discussion}

The $\sigma$ Ori E stars present a mystery for stellar evolution.
B-stars should not have large convective zones, and thus are expected
to possess relatively weak magnetic fields -- but typically measured
field strengths for $\sigma$ Ori E analogs are $ \sim 10$kG and
higher, in apparent (and strong) contradiction of this theoretical
expectation.  Furthermore, most theories predict that young B stars
should spin down on a $ \sim 1$Myr timescale, but the $\sigma$ Ori E
star HR7355 appears to be $15-25$Myr old \citep{Riv2013, Mikulasek10}
and yet has a high rotational velocity.  While we cannot measure the
magnetic field from our current data, the unique RRM signature clearly
indicates that HD 345439 and HD 23478 are also magnetized stars (which
future spectropolarimetry observations could confirm), and they are
both definitely fast rotators.  While the RRM can theoretically arise
in any star where the Alfven radius exceeds the Kelperian co-rotation
radius \citep{ud1, ud2}, in which case fields as low as $\sim 1$
kGauss could suffice, the other resemblances between these stars and
the $\sigma$ Ori E stars seem to imply that similar field strengths of
$\sim 10$ kGauss are most likely.  In the case of HD 23478, its sky
position, parallax of 4.99 mas, and proper motion of $+8 {\rm
  mas/yr}$, $-8 {\rm mas/yr}$ \citep{vanleeuwen07} all match the
members of the IC 348 young open cluster \citep{Scholz}.  This result
constrains the age of HD 23478 to match that of IC 348 -- previously
estimated as $1.3-3$ Myr by \citet{Herbig1998}; \citet{Bell2013}
however have recently derived an age closer to $\sim 5-6$ Myr based on
current isochrone-fitting techniques.  Future measurements of the
magnetic field in this star can then provide an estimated spindown
timescale, to see if this star matches expectations or, like HR 7355,
seems to be spinning too fast for its age and magnetic field.

As this work has shown, a particularly promising avenue for
identifying more of these unusual stars is near-infrared spectroscopy.
We believe the APOGEE spectra in Figure 1 to be the first published
NIR spectra of $\sigma$ Ori E stars, and they are remarkable in the
strength of the RRM signatures in the Brackett lines.  Based on our
sample of spectra (which are admittedly few in number and sparse in
phase sampling), each individual Brackett transition shows stronger
RRM signatures than \halpha for the same star, and the presence of
$10$ transitions in just $2/3$ of the H-band makes the NIR an
exceptionally powerful new diagnostic approach for identifying
$\sigma$ Ori E stars.  Our discoveries were entirely serendipitous,
yet they have increased the known sample of these stars by $\sim
10$\%, and HD 345439 alone has enhanced the number of ``extreme''
(near-breakup velocity) rotators by 50\%. Furthermore, the optical
spectra of these stars are much more ``normal'' than their NIR spectra
- both HD 23478 and HD 345439 have previous optical observations and
classifications that entirely missed their RRM nature.  We can see in
Figure 4 that the Brackett series RRM signatures are substantially
stronger than even the NIR Paschen series, indicating that the NIR
H-band may be a ``sweet spot'' for this diagnostic.  The fact that HD
23478 is both nearby and bright in the optical, yet eluded RRM
classification until now, further accentuates the diagnostic power of
NIR spectroscopy for this work.  Thus, with the advent of powerful IR
spectrographs at many observatories, and of large-scale IR
spectroscopic surveys such as APOGEE, we can speculate that the
discovery of these previously-rare stars may accelerate quickly in the
near future.

\acknowledgments

Funding for SDSS-III has been provided by the Alfred P. Sloan
Foundation, the Participating Institutions, the National Science
Foundation, and the U.S. Department of Energy Office of Science. The
SDSS-III web site is http://www.sdss3.org/.  DC and SRM gratefully
acknowledge support by National Science Foundation (NSF) grant AST11-
09718. The authors thank Kevin Covey for his helpful comments on the
IC 348 nature of HD 23478 and on the overall manuscript, and the anonymous referee for helpful comments.

SDSS-III is managed by the Astrophysical Research Consortium for the
Participating Institutions of the SDSS-III Collaboration including the
University of Arizona, the Brazilian Participation Group, Brookhaven
National Laboratory, University of Cambridge, Carnegie Mellon
University, University of Florida, the French Participation Group, the
German Participation Group, Harvard University, the Instituto de
Astrofisica de Canarias, the Michigan State/Notre Dame/JINA
Participation Group, Johns Hopkins University, Lawrence Berkeley
National Laboratory, Max Planck Institute for Astrophysics, Max Planck
Institute for Extraterrestrial Physics, New Mexico State University,
New York University, Ohio State University, Pennsylvania State
University, University of Portsmouth, Princeton University, the
Spanish Participation Group, University of Tokyo, University of Utah,
Vanderbilt University, University of Virginia, University of
Washington, and Yale University.

The Hobby-Eberly Telescope (HET) is a joint project of the University
of Texas at Austin, the Pennsylvania State University,
Ludwig-Maximilians-Universit\"{a}t M\"{u}nchen, and Georg-August-Universit\"{a}t
G\"{o}ttingen. The HET is named in honor of its principal benefactors,
William P. Hobby and Robert E. Eberly.

\end{document}